\documentclass[prb,twocolumn,draft,amsmath,showpacs]{revtex4}
\usepackage{graphics}

\begin{document}
    
\bibliographystyle{prsty}
\input epsf

\title {Diffusive energy transport in the $S=1$ Haldane chain compound 
AgVP$_2$S$_6$ }

\author {A.V. Sologubenko, S.M. Kazakov and  H.R. Ott}
\affiliation{Laboratorium f\"ur Festk\"orperphysik, ETH H\"onggerberg,
CH-8093 Z\"urich, Switzerland}
\author {T. Asano and Y. Ajiro}
\affiliation{Department of Physics, Kyushu University, Fukuoka 812-8581, Japan}

\date{\today}

\begin{abstract}
We present the results of measurements of the thermal conductivity $\kappa$ 
of the spin $S=1$ 
chain compound AgVP$_2$S$_6$ in the temperature range between  2 and 
300~K and with the 
heat flow directed either along or perpendicular to the chain direction.  
The analysis of the anisotropy of the heat transport allowed for the 
identification of 
a small but non-negligible magnon contribution $\kappa_{m}$ along the chains, 
superimposed on the dominant phonon contribution $\kappa_{\rm ph}$. 
At temperatures above about 100~K
the energy diffusion constant $D_{E}(T)$, calculated from the 
$\kappa_{m}(T)$ data, exhibits similar features as the spin diffusion constant 
$D_{S}(T)$, previously measured by 
NMR.  In this regime, the behaviour of 
both transport parameters is consistent with a diffusion process that 
is caused by 
interactions inherent to one-dimensional $S=1$ spin systems.
\end{abstract}
\maketitle

\section{Introduction}
Anomalous features of transport properties of 
one-dimensional Heisenberg antiferromagnetic (1D HAFM) spin $S=1/2$ systems 
were predicted decades ago. 
In particular, it was argued that both the spin and thermal conductivity 
are not 
expected to be of diffusive character as is typical for classical 
systems without long-range order,\cite{Kawasaki63,Stern65} but 
instead are based on 
ballistic transport of spin and energy, 
at least in the case of ideal 1D systems.\cite{Huber69,Huber69a,Niemeijer71,Krueger71}
On theoretical grounds it was recently demonstrated that this difference is 
due to the conservation of spin and energy currents in 
integrable models which apply for 1D S=1/2 spin 
systems.\cite{Castella95,Zotos97,Saito96,Narozhny98,Naef98} 
The situation is much less clear with respect to the theory of transport 
properties of 1D S=1 systems. 
Contrary to half-integer spin 1D HAFM systems with isotropic exchange, 
for which the spin 
excitations are $S=1/2$ spinons and the corresponding spectrum is gapless,\cite{Faddeev81}
the integer-spin chains exhibit gapped excitation spectra and $S=1$ magnon 
excitations.\cite{Haldane83} 
Numerical calculations for $S=1$, 1D HAFM systems relate the gap 
$\Delta$ to the exchange integral $J$, such that $\Delta=0.41 J$.\cite{Sorensen93}
Early theoretical work predicted that 
the transport in ideal 1D HAFM S=1 systems  must be diffusive.\cite{Huber69} 
More recent discussions 
of Sachdev and Damle,\cite{Sachdev97,Damle98} considering  the spin transport in Heisenberg 
S=1 chains in terms of the nonlinear $\sigma$-model (NL$\sigma$M), 
also concluded with the claim of diffusive spin transport. However, Fujimoto,\cite{Fujimoto99}
based on the integrability of NL$\sigma$M, suggested 
that in a perfect 1D S=1 system, the spin transport is ballistic. The author 
noted, however, that  external perturbations, unavoidable in real S=1 
chain compounds and  
leading to the destruction of integrability, may restore diffusive 
transport. In this respect, Fujimoto refers to a previous experimental observation of spin diffusion 
in the S=1  
chain compound AgVP$_{2}$S$_{6}$, by Takigawa {\em et al.},\cite{Takigawa96} 
and suggests that the spin-phonon 
interaction might provide this type of perturbation.  

Energy transport via spin excitations, which can be probed by 
measurements of the thermal 
conductivity, has not yet been investigated in S=1 chain compounds. 
It seems natural to expect that its  features are similar to those of spin 
transport. In a recent paper by Alvarez and Gros,\cite{Alvarez02_Con} relations between 
the spin and thermal conductivities in 1D spin systems, similar to the famous 
Wiedemann-Franz law for the ratio between the electrical and thermal 
conductivities in 
metals, were discussed. In the present work, we report the results of 
an experimental 
investigation of the thermal conductivity of  AgVP$_{2}$S$_{6}$, a compound 
which is considered as one of the best physical realizations of an $S=1$ 1D AFM Heisenberg 
model system.  AgVP$_{2}$S$_{6}$ crystallizes with a monoclinic crystal structure of type $P2/a$.\cite{Lee86}
Each $V^{3+}$ ion with spin $S=1$ is located 
inside a distorted octahedron of sulfur ions. The zigzag chains formed by 
these vanadium ions run along the $a$ axis. Evidence for an energy gap 
in the spin excitation spectrum was found by 
measurements of the magnetic susceptibility\cite{Colombet87,Payen90_Pow,Asano94,Mutka95_Fin} 
and inelastic neutron-scattering  experiments.\cite{Mutka89,Mutka91} 
The magnetic properties of the compound are well described by the 
isotropic nearest-neighbor Heisenberg Hamiltonian and weak 
single-ion anisotropy 
\begin{equation}
H = J \sum_{i} {\bf S}^{i} {\bf S}^{i+1} + D\sum_{i} (S_{z}^{i})^{2},
\label{eHamiltonian}
\end{equation}
with the intrachain exchange constant $J/k_{B}=780$~K and 
$D/J=5.8 \times 10^{-3}$.\cite{Takigawa95} The interchain 
interaction $J'\leq 10^{-5}J$ is very weak.\cite{Mutka91}
The dynamics of the V$^{3+}$ spins in AgVP$_{2}$S$_{6}$ was studied by NMR experiments by 
Takigawa et al.\cite{Takigawa96} 
The data are compatible with spin diffusion at 
temperatures exceeding $\sim 100 {\rm ~K}$ and above a 
temperature-dependent frequency of the order of 
$10^{11}-10^{12} {\rm ~s}^{-1}$. Using these results, 
the temperature dependence of the spin diffusion constant $D_{S}(T)$ 
was established. 
Below we compare the experimental results for the temperature variation of the spin diffusion 
constant $D_{S}(T)$, probed by NMR, and the energy diffusion 
parameter $D_{E}(T)$, which is obtained from the results of our thermal conductivity measurements. 

\section{Samples and experiment}

For the measurements of the thermal conductivity $\kappa(T)$, we grew several single crystals of  
AgVP$_2$S$_6$ by solid state reaction of stoichiometric amounts of  
Ag, V, P, and S, as described in
Ref.~\onlinecite{Lee86}. 
The largest crystals had dimensions 
of several mm along the $a$-axis (the spin chain direction), a maximum 
of
0.5~mm along the $b$-axis and, at most, 0.2~mm along the $c$-axis. A single 
crystalline specimen with dimensions of $2\times 0.5\times 0.15$~mm$^{3}$ (denoted as S1) was used for the 
thermal conductivity measurements along the chain direction. 
Another sample from the same batch served to measure 
the dc magnetic susceptibility $\chi$. 

A standard method for separating the 
spin-mediated thermal conductivity $\kappa_{m}(T)$ from other, less 
anisotropic, 
contributions to $\kappa(T)$ is based on measurements of this quantity both along 
and perpendicular to the chain direction. Unfortunately, the 
dimensions of our crystals only allowed for measurements of 
$\kappa(T)$ along the chains. 
For this reason we investigated the heat transport also on 
grain-aligned specimens. 
Some of these samples were previously used for measurements of the magnetic 
susceptibility\cite{Asano94} and for NMR experiments.\cite{Takigawa95,Takigawa96}  
Each of the specimens with typical sizes of $4
\times 2 \times 0.2 {\rm ~mm}^{3}$ were single pieces consisting of many equally oriented flake-like single 
crystals. On one such piece (denoted as sample J1), the thermal 
conductivity was measured along the $a$-axis.  Another 
sample, denoted as J2, with the largest spatial extension along the $b$-axis,  was 
cut out from the central 
part of sample J1. Sample J2 was subsequently used for experiments 
with the  heat flow perpendicular to the chain direction.

The thermal conductivity was measured using a standard uniaxial heat 
flow method, where the constant heat flux along a  rectangular-bar shaped 
sample was produced by a heater attached to one end of the sample. 
The opposite end of the sample was attached to a copper heat 
sink. The temperature gradient was monitored by a system of 25~$\mu$m diameter Chromel-Au+0.07\%Fe
thermocouples. We estimate the uncertainty of the absolute
value of $\kappa$ to be of the order of 10\%,  
because of the uncertainty of the
sample geometry.  The relative error of the measured thermal 
conductivity, however, is only about 0.5\%. 
The magnetic susceptibility was measured between 4.5 and 240~K, employing a 
commercial SQUID magnetometer.

\section{Results and discussion}

\begin{figure}[t]
 \begin{center}
  \leavevmode
  \epsfxsize=0.80\columnwidth \epsfbox {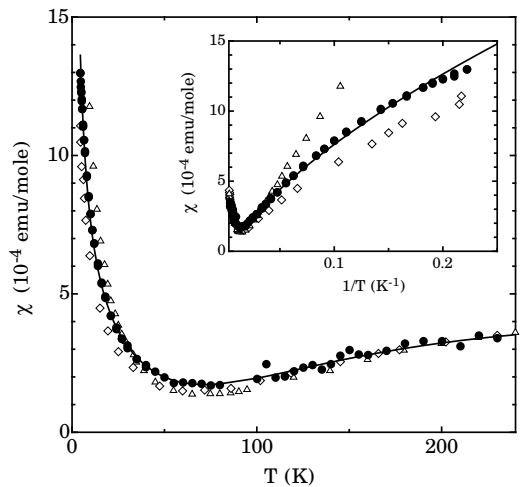}
   \caption{
  Magnetic susceptibility $\chi$ as a function of temperature.  
  Our results are represented by the solid circles. For comparison, 
  previous
  data reported in Ref.~\onlinecite{Mutka95_Fin} and obtained for a powder and a 
  single-crystalline sample are displayed by the open diamonds  and triangles, 
  respectively.
  }
\label{Chi}
\end{center}
\end{figure}

The  magnetic susceptibility data are shown in 
Fig.~\ref{Chi}.  The measurements were taken in a magnetic field of 
$H=40 {\rm ~kOe}$ oriented along the $a$-axis. The data are in good agreement 
with results of  Mutka {\em et al.},\cite{Mutka95_Fin} also shown in Fig.~\ref{Chi}. 
In particular, 
the nonlinear $\chi$ vs. $1/T$ dependence at low temperatures, noticed in 
Ref.~\onlinecite{Mutka93} and attributed to the weak coupling between finite-length 
segments of spin chains, is well reproduced in 
our data (see the inset of Fig.~\ref{Chi}). 
The $\chi(T)$ data can be fit to the equation\cite{Mutka95_Fin}
\begin{equation}
 \chi(T) = \frac{A}{T^{1/2}} \exp 
 \left(-\frac{\Delta}{k_{B}T}\right) + \frac{B}{T^{\alpha}} + \chi_{0},
\label{eChiT}
\end{equation}
where the first term on the right-hand side represents the contribution from the 
Haldane-gapped $S=1$ chains, the second term reflects the above 
mentioned finite segment interaction contribution and the last and
constant term is  due to the  diamagnetic orbital susceptibility. The 
best fit, shown as the solid line in Fig.~\ref{Chi}, is achieved with the fit parameters $A= 1.77 \times 10^{-2} {\rm 
~emu~mol^{-1}~K^{1/2}}$, $\Delta/k_{B}= 244 {\rm ~K}$, $B= 3.60\times 10^{-3} {\rm 
~emu~mol^{-1}~K^{0.55}}$, $\alpha = 0.55$, and $\chi_{0} = -2.5 \times 10^{-4 }{\rm ~emu~ 
mol^{-1}}$. The 
resulting value for the energy gap $\Delta/k_{B}$ is consistent with 
estimates of 228~K (Ref.~\onlinecite{Mutka95_Fin}) and 250~K 
(Ref.~\onlinecite{Asano94})
from the analyses of magnetic susceptibilities, but somewhat lower than the 
results obtained from NMR\cite{Takigawa95} (320~K) and 
the neutron scattering\cite{Mutka91} (300~K) measurements. The 
power-law exponent $\alpha=0.55$ is within the range $0.5 < \alpha < 
0.8$ established in Ref.~\onlinecite{Mutka95_Fin}.

\begin{figure}[t]
 \begin{center}
  \leavevmode
  \epsfxsize=1\columnwidth \epsfbox {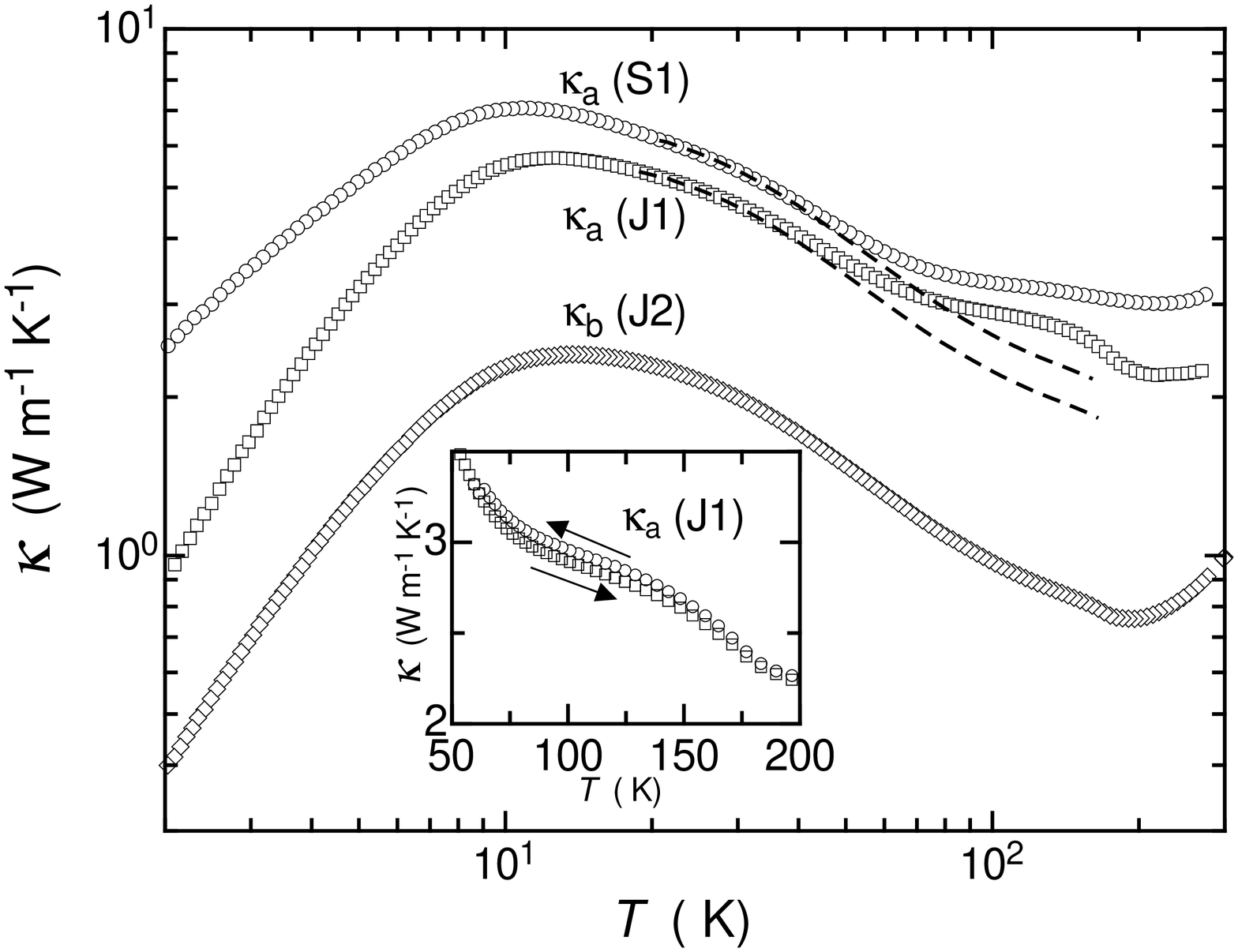}
   \caption{
  Thermal conductivity vs. temperature for two samples along the 
  chain direction and one sample in a perpendicular direction. The 
  inset demonstrates the slightly irreversible behaviour at high 
  temperatures.
  }
\label{KappaAll}
\end{center}
\end{figure}

The temperature dependences of the different thermal conductivities are shown in 
Fig.~\ref{KappaAll}. Each curve exhibits a maximum at temperatures 
around 10~K and a region with a positive slope $\partial \kappa 
/\partial T$ at temperatures above about 200~K. 
The latter feature is presumably 
due to unaccounted heat losses via radiation, typical in 
standard steady-state thermal conductivity experiments probing 
materials with 
low values of $\kappa$ at high temperatures. 
Our numerical  estimates suggest that for our AgVP$_2$S$_6$ samples, 
these effects are negligible only below about 170 to 200~K and 
therefore, our data  
analysis presented below is restricted to temperatures below 170~K. 

By rapidly reducing the temperature of the grain-aligned samples (J1 and J2) 
from values in the region between 170 and 300~K to temperatures between 
80 and 170~K with a cooling rate of approximately 1~K/s, subsequent slow 
relaxations of $\kappa$ with time were observed.  
As illustrated in the inset of Fig.~\ref{KappaAll},
even upon a slow variation of the temperature (as slow as 
0.01~K/s), a small hysteresis of $\kappa(T)$ was observed at 
intermediate temperatures above the maximum of $\kappa(T)$. 
We believe that these effects are not related to intrinsic properties of 
the material. Since they were not observed when probing the single-crystalline 
specimen, they are most likely due to the multigrain nature of these samples. 

By inspecting the $\kappa(T)$ curves at temperatures below 60~K, we 
note that they all exhibit the same qualitative features.
At higher temperatures, 
a shoulder-type feature in $\kappa_{a}(T)$, which is absent 
in the $\kappa_{b}(T)$ curve, may be identified. Similar high-temperature features
emerging as a shoulder or a second maximum were observed for $\kappa(T)$ 
if measured along the chain or ladder direction of spin-chain and 
spin-ladder compounds. 
\cite{Salce98,Ando98,Takeya00,Sologubenko00_Lad,Hess01,Kudo01_Spi,Sologubenko00_213,Sologubenko01} 
An obvious interpretation of this 
high-temperature feature is to ascribe it to a spin-mediated thermal conductivity 
$\kappa_{m}$, in addition to  the common heat conduction via phonons, 
which is
reflected in a contribution $\kappa_{\rm ph}$ to the total thermal 
conductivity.

In order to isolate the spin contribution $\kappa_{m}(T)$, accurate 
evaluations of other contributions to the total measured 
$\kappa(T)$ need to be made. Since AgVP$_2$S$_6$ is an insulator, no 
heat conduction by free 
charge carriers is expected, and only heat 
transport via lattice excitations needs to be considered. Because of the extremely weak 
interaction between spins of neighboring chains, no sizable heat 
transport  via the spin system is expected for directions perpendicular to the 
$a$-axis. Hence only phonons contribute to $\kappa_{b}(T)$, 
shown in Fig.~\ref{KappaAll}. The shape of the $\kappa_{b}(T)$ curve,  typical 
for phonon heat transport, results from the competition between different 
phonon scattering mechanisms, 
most notably phonon scattering by sample or grain boundaries at the 
lowest temperatures and by phonons, defects and magnons at more 
elevated 
temperatures. In contrast to the boundary and defect scattering which is usually sample-dependent, 
the scattering of phonons by quasiparticles in general is expected to be sample-independent 
because it is related to intrinsic scattering processes.

At temperatures much less than  $\Delta/k_{B}$, the number of spin 
excitations that participate in the heat transport 
reflected in $\kappa_{a}(T)$ is decreasing 
exponentially with decreasing temperature. 
Therefore, at $T \ll \Delta/k_{B}$ the phonon heat 
transport is expected to also dominate $\kappa_{a}(T)$. 
While the different temperature dependences of 
$\kappa$ for different samples on the left of the $\kappa(T)$ maxima 
in Fig.~\ref{KappaAll} can be accounted for by 
differences in the boundary scattering conditions, it is remarkable 
that at 
temperatures above the maxima, up to about 40~K, 
the ratio between the
$\kappa$ values for different samples is practically independent of
temperature. This implies that the temperature dependence of these 
$\kappa(T)$ curves are identical and  that in this temperature region, the 
intrinsic processes of phonon scattering dominate. As we have 
already argued in Ref.~\onlinecite{Sologubenko01}, the anisotropy of 
the purely 
phononic heat transport is not likely to change at high temperatures. 
This is why we assume that at high temperatures, the phonon thermal 
conductivity along different directions and for different 
samples merely differs by a constant factor. 
With this assumption, we calculated the magnon 
contribution 
\begin{equation}
\kappa_{m,a} = \kappa_{a} - K*\kappa_{b}
\label{eKmcalc}
\end{equation}
with $K=2.7$  and  $K=2.3$ for samples S1 and 
J1, respectively. The $K$ values were calculated as the averaged ratio $\kappa_a/\kappa_b$ 
in the temperature interval between about 20 and 35~K.
The  phonon contributions $K*\kappa_{b}$ for 
samples S1 and J1 are shown by the broken lines in Fig.~\ref{KappaAll}.

\begin{figure}[t]
 \begin{center}
  \leavevmode
  \epsfxsize=1\columnwidth \epsfbox {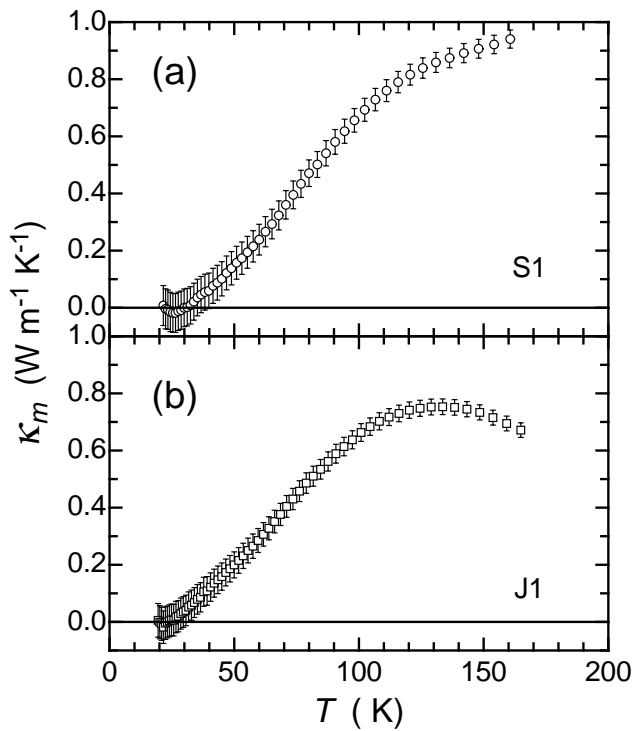}
   \caption{
  Magnon thermal conductivity along the $a$-axis of AgVP$_2$S$_6$, extracted from 
  $\kappa_{a}(T)$ for the two samples S1 (a) and J1 (b). 
  }
\label{KappaS}
\end{center}
\end{figure}

The resulting values of $\kappa_{m}(T)$ are presented in Fig.~\ref{KappaS}.
The uncertainties caused by the subtraction of two experimentally measured curves, 
each with a relative error of about 0.5\%, are also displayed in Fig.~\ref{KappaS}.  
Because of the relatively large uncertainties inherent to the 
evaluation procedure below about 40~K, it is impossible 
to draw any conclusions about $\kappa_{m}(T)$  in that temperature 
region. Therefore, only data at temperatures above 50~K were 
considered for the following analysis. 
In the accessible regime, $\kappa_{m}(T)$ first increases with 
increasing temperature but tends to pass through a maximum or to 
reach saturation at higher temperatures. 
The maximum absolute values of  $\sim 1$~W~m$^{-1}$~K$^{-1}$ for the 
magnon thermal conductivity 
are one to two orders of magnitude smaller than the corresponding 
values for $\kappa_{m}$ in $S=1/2$ two-leg spin-ladder 
compounds (La,Sr,Ca)$_{14}$Cu$_{24}$O$_{41}$ 
\cite{Sologubenko00_Lad,Hess01,Kudo01_Spi} 
and various $S=1/2$ spin-chain materials, 
\cite{Salce98,Takeya00,Sologubenko00_213,Sologubenko01,Sologubenko03_Uni} where the 
enhanced spin-mediated thermal conductivity is thought to be the 
consequence of a quasiballistic energy transport. 

\begin{figure}[t]
 \begin{center}
  \leavevmode
  \epsfxsize=1\columnwidth \epsfbox {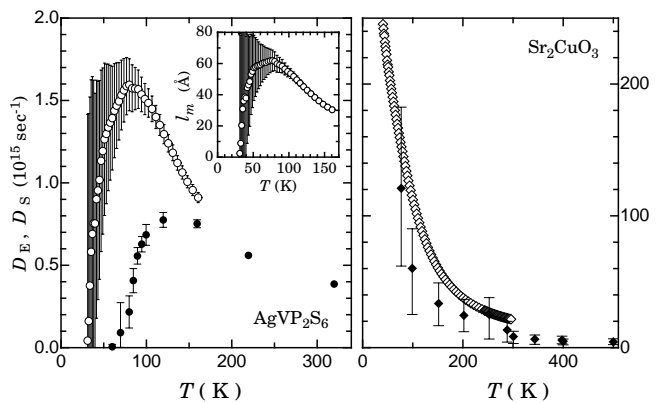}
   \caption{
  The energy diffusion constant $D_{E}(T)$, calculated from the thermal 
  conductivity data of sample S1 (open circles), and the spin diffusion constant 
  $D_{S}(T)$ from Ref.\onlinecite{Takigawa96} (solid circles) are 
  shown in the left panel. In the right panel, the corresponding data for 
  the $S=1/2$ spin chain compound Sr$_2$CuO$_3$ are shown. 
  }
\label{DeDscomp}
\end{center}
\end{figure}

In view of the implicit interpretation that heat may be transported 
via spin excitations, it seems natural to compare the energy transport with spin 
transport. The temperature dependence of the spin diffusion constant 
$D_{S}(T)$ in AgVP$_2$S$_6$ was measured by Takigawa {\em et al.}\cite{Takigawa96}
Based on our $\kappa_{m}(T)$ data, we calculated the 
spin-related energy diffusion constant $D_{E}$ from  
$D_{E}(T) = \kappa_{m}(T) / (C_{s}(T) a^{2})$, 
where $C_{s}(T)$ is the 
specific heat of the spin system and $a=2.96 {\rm ~\AA}$ is the 
distance between neighboring spins along the chains. The 
low-temperature specific heat 
of Heisenberg type $S=1$ spin chains
was analyzed by Jolicoeur and Golinelli
\cite{Jolicoeur94} employing the quantum nonlinear $\sigma$-model. 
The energy $\varepsilon$ versus the wavevector $k$, measured 
from the AFM wave vector $\pi/a$, is given by the equation 
\begin{equation}
\varepsilon(k) = \left[ V^{2}(ka)^{2} + \Delta^{2}  \right]^{1/2},
\label{eDisp}
\end{equation}
and the results of numerical calculations at  $T=0$ are 
$\Delta_0 = 0.41 J$ and $V = 2.49 J$.\cite{Sorensen93} The energy 
diffusion constant $D_{E}(T)$, calculated for 
the single-crystalline sample S1 as outlined above and using $C_{s}(T)$ 
as calculated by Jolicoeur and Golinelli, 
\cite{Jolicoeur94,JolicoeurPriv} is shown in the left 
panel of Fig.~\ref{DeDscomp}; the data for the second sample, not shown here, are very similar. 
The spin diffusion constant $D_{S}(T)$ from 
Ref.~\onlinecite{Takigawa96} is also displayed in the same figure. 
The absolute values of the two parameters are similar in magnitude  
at high temperatures and they exhibit very similarly shaped temperature dependences. 
This is true although, in comparison with the corresponding feature of 
$D_{S}(T)$, the maximum value of $D_{E}(T)$ is about two times 
larger and slightly shifted to lower 
temperatures. 
The only other spin-chain compound for which data on both 
the spin and the energy transport are available, is the 
Heisenberg $S=1/2$ chain compound Sr$_{2}$CuO$_{3}$. 
In the right panel of Fig.~\ref{DeDscomp}, we present both $D_{S}(T)$ data from 
Ref.~\onlinecite{Thurber01} and $D_{E}(T)$ data calculated from results of 
our
previous thermal conductivity measurements on Sr$_{2}$CuO$_{3}. 
$\cite{Sologubenko01}
It is again obvious that both quantities vary similarly with temperature 
and $D_{E} \sim 2 D_{S}$ in the covered temperature interval.
The only other material for which
$D_{E}$ and $D_{S}$ have been compared up to now is solid
$^{3}$He in the region $0.05 < T < 0.12 {\rm ~K}$, where the ratio
$D_{E}/D_{S} \approx 2.1$  is, again, close to two.\cite{Hunt68} It 
is  remarkable that indeed the factor of 2 difference between $D_{E}$ and 
$D_{S}$ is expected to be a general property of any Bravais lattice 
with identical spins on each lattice point.
\cite{Redfield68}  The fair agreement of our results with this 
expectation thus confirms the reliability of our calculation of 
the spin-mediated thermal conductivity. 

The limited accessible temperature interval  for the evaluation of 
$\kappa_{m}(T)$ and the increasing uncertainty with decreasing temperatures do not allow a serious 
analysis of its overall temperature dependence. The classical high-temperature 
limit of $D_{E}$ for 
a  spin system is given by\cite{Huber69}
\begin{equation}
D_{E,{\rm ht}} \sim J [S (S+1)]^{1/2}/\hbar.
\label{eDeHT}
\end{equation}
For AgVP$_2$S$_6$, this 
high-temperature limit is    $D_{E,{\rm ht}} \sim  10^{14} 
{\rm ~sec}^{-1}$. The observed values in the temperature range 
covered in this study are, except at the lowest temperatures, consistently higher than  $D_{E,{\rm 
ht}}$, but tend to approach this value with increasing temperature. 
The enhancement is, however, much less pronounced than in 
$S=1/2$ chain and ladder compounds. A better way to demonstrate this is 
by invoking the average mean free path 
of itinerant  spin excitations $\ell_{m}(T)$, which can be defined as $\ell_{m} \equiv 
D_{E} / v_{m} $, where $v_{m}(T)$ is the average group velocity 
\begin{equation}\label{Vs_ave}
  v_{m}(T)= {\frac{1}{\hbar}} \left[ 
  \int {\frac{\partial\varepsilon}{\partial k}}  f(\varepsilon,T) dk  \right] 
  / \left[ \int   f(\varepsilon,T) dk
  \right],
\end{equation}
with $f(\varepsilon,T) = (\exp (\varepsilon /k_B T) - 1)^{-1}$ and  
the dispersion relation 
$\varepsilon(k)$ as given by Eq.~\ref{eDisp}. 
We calculated $\ell_{m}(T)$  taking into account the temperature 
dependence of the energy gap, given by the NL$\sigma$M at low temperatures as\cite{Jolicoeur94} 
\begin{equation}
\Delta(T) \approx \Delta_0 + (2\pi \Delta_0 k_{B}T )^{1/2} 
\exp(-\Delta_0 /k_{B}T).
\label{eDeltavsT}
\end{equation}
The calculated values of $\ell_{m}(T)$, shown in the inset of Fig.~\ref{DeDscomp},
do not exceed 60~\AA. In contrast, the mean free paths of spin 
excitations in many $S=1/2$ chains typically reach values of the order of 
10$^{3}$~\AA.\cite{Salce98,Ando98,Takeya00,Sologubenko01,Sologubenko03_Uni}

The data presented above and the subsequent comparison of $S=1$ and 
$S=1/2$ spin chains provoke the obvious question concerning the 
reasons for the much 
reduced spin-mediated energy transport in the $S=1$ system. 
Two possibilities may be 
considered. First, extrinsic causes, such as a much stronger scattering of magnons by 
defects and phonons might significantly reduce $\kappa_{m}$ in the 
compound studied here. Second, intrinsic causes of reduced energy 
transport, e.g., as a result of the non-integrability of the Hamiltonian even 
for the pure $S=1$ spin system should be considered. 
We argue that the influence of defects is 
hardly the major factor leading to low values of $D_{E}$, 
because the data for the two different samples of  different quality 
that were studied in the course of this work provide  
very similar results for the magnon thermal conductivity (see 
Fig.~\ref{KappaS}). That the magnon-phonon scattering is 
the main source for low $\kappa_{m}$ values in AgVP$_2$S$_6$ cannot 
altogether be ruled out, but at present we cannot offer an obvious reason 
why this scattering should be much stronger in AgVP$_2$S$_6$ than in 
all other spin-chain compounds mentioned above. 
 
The second and most likely possibility is that the energy diffusion 
observed in our experiments on AgVP$_2$S$_6$ is intrinsic and is governed by dynamic 
correlations  of the $S=1$ spin system.  Unfortunately, to our knowledge, there is no 
calculation of $D_{E}(T)$ available in the literature for 
temperatures much less than $J/k_{B}$ to compare our data with. 
However, a semiclassical consideration of 
$S=1$ chains by Damle and Sachdev\cite{Damle98} and applied to 
AgVP$_2$S$_6$, results in $D_{S}$ values which are in fair agreement 
with the $D_{S}(T)$ data shown in Fig.~\ref{DeDscomp}.  Since our 
values for $D_{E}$ are compatible with $D_{S}$, 
we take this as a clear indication for the same type of behavior of both 
parameters. 
Nevertheless, the reason for the divergence between $D_{E}(T)$ 
and $D_{S}(T)$ at low temperatures, namely below about 100~K is not understood. In 
Ref.~\onlinecite{Takigawa96}, the drop of $D_{S}(T)$ at temperatures below 
100~K was considered to be the result of a crossover 
from spin diffusion to a region of freely propagating magnons. Our 
results do not contradict this assumption, however a rigorous analysis of 
$D_{E}(T)$ in this temperature regime is impossible, because of the 
prohibitively large uncertainty 
in $\kappa_{m}$ at these 
temperatures.

A rigorous approach to transport properties of low-dimensional 
quantum spin systems,
based on the notion of 
a thermal Drude weight $D_{\rm th}(T)$ in integrable spin systems, 
analogous to the familiar Drude weight in the 
theory of metallic conductivity,  has recently  been developed. 
\cite{Kluemper02,Alvarez02_Low,Alvarez02_Ano,Alvarez02_Con,Saito02,HeidrichMeisner02,Orignac03,Saito03}
The spin-related thermal conductivity $\kappa_{\rm spin}$, which here
corresponds to a conductivity via either magnons in the $S=1$ case ($\kappa_{m}$) or 
spinons in the $S=1/2$ case ($\kappa_{s}$),   is  given as  
\begin{equation}
\kappa_{\rm spin}(T)=\lim_{\omega\to 0} \kappa_{\rm spin} 
(\omega ) 
\label{eDth}
\end{equation}
with
\begin{equation}
    \kappa_{\rm spin}(\omega ) = D_{\rm th}(T) \delta(\omega).
\label{eDth2}
\end{equation}
For an ideal integrable system,  $\kappa_{\rm spin}$ diverges if $D_{\rm th}$ is nonzero. 
If perturbations introduce a finite lifetime $\tau$, the thermal 
conductivity is also expected to be finite and can be represented as $\kappa_{\rm 
spin} = D_{\rm th}\tau$. 
In Ref.~\onlinecite{Orignac03}, where this approach was applied to S=1 
HAFM chains, it is noted that this is true only 
in the case of a large and energy-independent $\tau$ or, 
equivalently, to a  long mean free path  $\ell$. However, impurity scattering, 
for example,  
is shown\cite{Orignac03} to invoke energy-dependent lifetimes, thus 
leading to a more 
complicated behaviour of $\kappa_{m}$ which is not simply given by a product of the thermal 
Drude weight and the average relaxation time.  
Since in our case, as reflected in the small values of 
$\ell_{m}$ which approach the interspin distance (see the inset of Fig.~\ref{DeDscomp}),
the relaxation is quite strong, we do not attempt a direct 
application of the thermal Drude weight formalism to our result for 
$\kappa_{m}(T)$ of
AgVP$_{2}$S$_{6}$. This approach seems, however, well justified for 
the analysis 
of $\kappa_{s}$ in $S=1/2$ 1D spin systems, where the spinon mean 
free paths are often rather large.\cite{Sologubenko03_Uni} 

\section{Summary and conclusions}

In conclusion, by probing the anisotropy of the thermal conductivity of 
the S=1 Haldane-gap HAFM compound AgVP$_{2}$S$_{6}$, we established 
the magnitude and the temperature variation of the spin-related energy transport in the temperature 
region between about 50 and 170~K. 
By comparing the energy and the spin diffusion parameters $D_{E}(T)$ 
and $D_{S}(T)$, both derived from experimental data, we note that 
they not only exhibit similar temperature dependences but that also 
their absolute values are of the same order of magnitude.
We argue that the previous suggestion of an
intrinsic origin of the spin diffusion at high temperatures in this 
system\cite{Damle98} also applies to
the energy transport that is investigated here. The character of the energy 
transport at low temperatures remains to be investigated.

\acknowledgments
This work was financially supported in part by
the Schweizerische Nationalfonds zur F\"{o}rderung der Wissenschaftlichen
Forschung.

\end{document}